\documentclass[%
 reprint,
 amsmath,amssymb,
 aps,
]{revtex4-2}

\usepackage{graphicx}
\usepackage{dcolumn}
\usepackage{bm}
\usepackage{amsmath}
\usepackage{slashed}
\usepackage{revsymb}

\makeatletter
\newcommand\makebig[2]{%
  \@xp\newcommand\@xp*\csname#1\endcsname{\bBigg@{#2}}%
  \@xp\newcommand\@xp*\csname#1l\endcsname{\@xp\mathopen\csname#1\endcsname}%
  \@xp\newcommand\@xp*\csname#1r\endcsname{\@xp\mathclose\csname#1\endcsname}%
}
\makeatother

\makebig{biggg} {4.0}
\makebig{Biggg} {4.5}
\makebig{bigggg}{5.0}
\makebig{Bigggg}{5.5}

\begin{document}

\preprint{APS/123-QED}

\title{Evanescent Electron Wave Spin}

\author{Ju Gao}
\email{jugao2007@gmail.com}
\author{Fang Shen}
\affiliation{University of Illinois, Department of Electrical and Computer Engineering, Urbana, 61801, USA}




\date{\today}

\begin{abstract}
This study demonstrates the existence of an evanescent electron wave outside both finite and infinite quantum wells by solving the Dirac equation and ensuring the continuity of the spinor wavefunction at the boundaries. We show that this evanescent wave shares the spin characteristics of the wave confined within the well, as indicated by analytical expressions for the current density across all regions. Our findings suggest that the electron cannot be confined to a mathematical singularity and that quantum information, or quantum entropy, can leak through any confinement. These results emphasize that the electron wave, fully characterized by Lorentz-invariant charge and current densities, should be considered the true and sole entity of the electron.

\end{abstract}


\maketitle


\section{\label{sec:wave}Electron wave spin}
Electron spin, a fundamental property of electrons, plays a crucial role in advancing fields such as quantum computing and spintronics~\citep{QComputing2019, QComputing2013, Spintronics2020}. Despite its significance, the nature and origin of electron spin have been debated since its discovery over a century ago~\citep{Stern22, Uhlenbeck26}. One of the main challenges is reconciling the concept of the electron as a spinning particle with the principles of special relativity. Additionally, the true radius of the electron’s corpuscular charge ball or disk remains undetermined and unmeasured. This uncertainty has led many to view electron spin as an abstract property rooted in mathematical formalism rather than as a tangible physical entity.

However, alternative interpretations have been proposed~\citep{Belinfante39, Ohanian86} that relate electron spin to the physical quantity of current density, $\pmb{j}$. In these frameworks, spin emerges as a measurable wave property because current density can be derived as an observable from the wavefunctions of the Dirac equation. Recently, we demonstrated the existence of stable circulating current densities of an electron within quantum wells~\citep{GaoJOPCO22}. The electron spin value of $\hbar/2$ is obtained by integrating the interaction term, $\pmb{j} \cdot \pmb{A}$, between the current density and an external vector potential over the spatial extent of the wavefunctions. This spin value is further influenced by the geometry of the quantum well and the quantum numbers of the states, highlighting the wave nature of the electron spin. We have suggested that this effect can be quantitatively verified through abnormal Zeeman splitting experiments. Moreover, this electron wave spin exhibits geometric and topological features that are absent in the traditional spinning particle model but could also be experimentally observed~\citep{GaoQeios2023}. In summary, it is the electron wave that is physically spinning.

Importantly, the circulating current density and charge density, $\rho$, together form a Lorentz-invariant four-current, $(\rho,\pmb{j})$. If the electron wave behaves in accordance with special relativity, then the electron as a discrete corpuscular entity with a well-defined radius cannot spin nor exist. This perspective suggests that the electron wave itself is the fundamental reality of the electron. By interpreting the electron purely as a physical wave, without invoking wave-particle duality, we provide a framework to reconcile some of the most perplexing aspects of quantum mechanics. For instance, the phenomenon of an electron appearing in multiple locations simultaneously is naturally explained by its wave nature, while the correlation between two entangled electrons over a distance can be understood as the overlapping of two physical waves.

This wave-based perspective also has significant implications for practical applications in quantum devices, such as quantum wires, which have been extensively studied using Schrödinger's equations~\cite{Lima_2008, qasem2022}. Since a wave cannot be entirely confined within a device, it becomes intriguing to explore the properties of the electron wave in the vicinity of the device as part of the overall electron entity. In our previous work, we examined the electron wave spin within an infinite quantum well, where the wavefunction outside the well is typically assumed to be zero~\cite{QMJoachain2000}. However, in reality, all potentials are finite, resulting in a non-negligible wavefunction outside the well. Of particular interest is the situation where the electron's eigenenergy is below the quantum well potential, leading to an evanescent wave that does not propagate but remains near the quantum well. We aim to study the spin properties of this evanescent wave and its relationship to the confined wave within the quantum well. This could open up possibilities for probing spin properties externally, without collapsing the spin state within the well, analogous to evanescent wave sensing in optics~\citep{Evanescent2019, RefaieSR2023, RefaieEPJST2023}.

This paper is structured as follows: in Section~\ref{sec:Finite}, we rigorously solve the Dirac equation for an electron in a quantum well and derive analytical expressions for the corresponding wavefunctions. In Section~\ref{sec:Evanes}, we calculate the current densities both inside and outside the quantum well, demonstrating the existence of an evanescent wave even in the case of an infinite quantum well. In Section~\ref{sec:EvaneSpin}, we focus on the spin behavior of the evanescent wave and discuss the broader implications for quantum mechanics interpretations and technological applications.

\section{\label{sec:Finite}Dirac electron in a finite cylindrical quantum well}
Here, we derive exact solutions of the Dirac equation in a finite cylindrical quantum well and obtain analytical expressions for the current density. The eigenenergy values will be solved numerically through the boundary condition equations.

The Dirac equation in the cylindrical coordinate is
\begin{equation}\label{Dirac}
\frac{1}{c} \frac{\partial }{\partial t}\psi(\pmb r,t)=\{-\pmb{\alpha} \cdot \pmb{\nabla}
-i\frac{mc^2}{\hbar c}\gamma ^0-i\frac{1}{\hbar c}U(\pmb r)\}\psi(\pmb r,t),
\end{equation}
where $(\rho,\phi,z)$ represent polar, azimuthal angle and z coordinate, respectively. The operator 
\begin{equation}
\pmb{\alpha} \cdot \pmb{\nabla}
=\alpha _{\rho }\frac{\partial }{\partial \rho }+\alpha _{\phi }\frac{1}{\rho }
\frac{\partial }{\partial \phi }+\alpha _z
\frac{\partial }{\partial z}
\end{equation}
contains $\alpha-$matrix in cylindrical coordinate
\begin{eqnarray}
\alpha _{\rho }&=&
\left(
\begin{array}{cccc}
 0 & 0 & 0 & e^{-i \phi } \\
 0 & 0 & e^{i \phi } & 0 \\
 0 & e^{-i \phi } & 0 & 0 \\
 e^{i \phi } & 0 & 0 & 0 \\
\end{array}
\right); \nonumber \\
\alpha _{\phi }&=&
\left(
\begin{array}{cccc}
 0 & 0 & 0 & -i e^{-i \phi } \\
 0 & 0 & i e^{i \phi } & 0 \\
 0 & -i e^{-i \phi } & 0 & 0 \\
 i e^{i \phi } & 0 & 0 & 0 \\
\end{array}
\right); \nonumber \\
\alpha _z &=&
\left(
\begin{array}{cccc}
 0 & 0 & 1 & 0 \\
 0 & 0 & 0 & -1 \\
 1 & 0 & 0 & 0 \\
 0 & -1 & 0 & 0 \\
\end{array}
\right),
\end{eqnarray}
with the following properties
\begin{eqnarray}
&&\sigma _{\rho }^2=\sigma _{\phi }^2=\sigma _{z}^2=1, \nonumber \\
&&\sigma _{\rho }\sigma _{\phi }=i \alpha _z, \nonumber \\
&&\sigma _{\rho }\sigma _{\phi }+\sigma _{\phi }\sigma _{\rho }=0. \nonumber \\
\end{eqnarray}

We now let the potential 
\begin{equation}
U(\pmb r)=U(\rho)=\Bigl\{
\begin{array}{cc}
 0, & \rho <R \\
 U, & \rho >R \\
\end{array}
\end{equation}
represent a finite cylindrical quantum well of potential $U$ and radius $R$. The wavefunction in the quantum well can be expressed by the separation of variables
\begin{equation}\label{psi}
\psi(\pmb r,t)=e^{-i \mathcal{E} t/\hbar}
e^{i P_z z/\hbar}\tilde{\psi }(\phi,\rho), 
\end{equation}
where the electron is not confined along the $z$-direction and behaves as a travelling wave. We further assume the wavelength of the travelling wave is much larger than the size of the quantum well, thus for our discussion, $P_z=0$. The eigenenergy, $\mathcal{E}$, is determined by the boundary conditions later. 

Plugging Eq.~\ref{psi} into Eq.~\ref{Dirac} to obtain the equation 
\begin{equation} \label{tildepsi}
\{-i\frac{\mathcal{E}}{\hbar c}+
\alpha _{\rho }\frac{\partial }{\partial \rho }+\alpha _{\phi }\frac{1}{\rho }\frac{\partial }{\partial \phi }+i\frac{mc^2}{\hbar c}\gamma ^0+i\frac{1}{\hbar c}U(\rho)\}\tilde{\psi }(\phi,\rho)=0. 
\end{equation}

$\tilde{\psi }(\phi,\rho)$ is a four-spinor that can be written as
\begin{equation}\label{tildepsiform}
\tilde{\psi }(\phi,\rho)=\left(
\begin{array}{c}
\mu_A(\phi,\rho) \\
\mu_B(\phi,\rho) \\
\end{array}
\right),
\end{equation}
where $ \mu_A(\phi ,\rho ) $ and $ \mu_B(\phi ,\rho )$ are two component spinor wavefunctions known as the large and small components of the Dirac wavefunctions that follow the equations
\begin{eqnarray}\label{muAmuB}
&&-i \frac{\mathcal{E}-U(\rho)-m c^2}{\hbar c} \mu_A(\phi ,\rho )=\nonumber \\
&&\left(
\begin{array}{cc}
 0 & e^{-i \phi }\frac{\partial }{\partial \rho }-i e^{-i \phi }\frac{1}{\rho }\frac{\partial }{\partial \phi } \\
 e^{i \phi }\frac{\partial }{\partial \rho }+i e^{i \phi }\frac{1}{\rho }\frac{\partial }{\partial \phi }  & 0 \\
\end{array}
\right)\mu_B(\phi ,\rho ); \nonumber \\
&&-i \frac{\mathcal{E}-U(\rho)+m c^2}{\hbar c} \mu_B(\phi ,\rho )=\nonumber \\
&&\left(
\begin{array}{cc}
 0 & e^{-i \phi }\frac{\partial }{\partial \rho }-i e^{-i \phi }\frac{1}{\rho }\frac{\partial }{\partial \phi } \\
 e^{i \phi }\frac{\partial }{\partial \rho }+i e^{i \phi }\frac{1}{\rho }\frac{\partial }{\partial \phi }  & 0 \\
\end{array}
\right)\mu_A(\phi ,\rho ). \nonumber \\
\end{eqnarray}

The above equations are combined to give the equation for $ \mu_A(\phi ,\rho ) $,
\begin{equation}\label{muA}
\left(\frac{\partial^2 }{\partial \rho^2 }+\frac{1}{\rho }\frac{\partial }{\partial \rho }+\frac{1}{\rho^2 }\frac{\partial^2 }{\partial \phi^2}\right)\mu_A(\phi ,\rho )=\Bigl\{
\begin{array}{cc}
-\zeta^2 \mu_A(\rho ), & \rho <R; \\\\
 \xi^2 \mu_A(\rho ),   & \rho >R,
\end{array}  
\end{equation}
where $\zeta$ and $\xi$ are wave numbers inside and outside the quantum well, respectively, 
\begin{eqnarray}\label{zetaxi}
\zeta &=&\sqrt{\frac{\mathcal{E}^2-m^2 c^4}{\hbar ^2 c^2}}, \nonumber \\
\xi &=&\sqrt{\frac{m^2 c^4 -(\mathcal{E}-U)^2}{\hbar ^2 c^2}},
\end{eqnarray}
and are both real numbers, since the quantum well potential falls in the range of
\begin{equation}
\mathcal{E}-m c^2<U<m c^2.
\end{equation}

We now separate the variables of $ \mu_A(\phi ,\rho ) $ for the spin-up electron, 
\begin{equation}\label{muAform}
\mu_A(\phi ,\rho )=e^{i l \phi }\mu_A(\rho )\left(
\begin{array}{c}
1 \\
0 \\
\end{array}
\right),
\end{equation}
where $l$ is the azimuthal quantum number for the angular wavefunction $ e^{i l \phi }  $. 

The radial wavefunction $ \mu_A(\rho ) $ follows the equation by plugging Eq.~\ref{muAform} into Eq.~\ref{muA},
\begin{equation}\label{muArho}
\left(
\frac{\partial^2 }{\partial \rho^2 }+\frac{1}{\rho }\frac{\partial }{\partial \rho }-\frac{l^2}{\rho^2 }
\right)\mu_A(\rho )=\Bigl\{
\begin{array}{cc}
-\zeta^2 \mu_A(\rho ), & \rho <R; \\\\
 \xi^2 \mu_A(\rho ),   & \rho >R.  
\end{array}  
\end{equation}

$ \mu_A(\rho ) $ can be readily solved from Eq.~\ref{muArho} and 
$ \mu_B(\phi ,\rho ) $ can be subsequently obtained from Eqs.~\ref{muAmuB}. Finally, the four component spinor wavefunction $\tilde{\psi }(\phi,\rho)$ in Eq.~\ref{tildepsiform} for the spin up electron is obtained 
\begin{widetext}
\begin{equation} \label{tildepsi2}
\tilde{\psi }(\phi,\rho)=\Bigggg \{
\begin{array}{ll}
e^{i l \phi }
\left(
\begin{array}{c}
 J_l(\zeta \rho ) \\
 0 \\
 0 \\
i e^{i \phi } \frac{\hbar c}{\mathcal{E}+m c^2} \{\frac{1}{2} \zeta \left[J_{l-1}(\zeta \rho )-J_{l+1}(\zeta \rho ) \right] -\frac{l}{\rho} J_{l}(\zeta \rho ) \} \\
\end{array}
\right), & \rho \leq R; \\\\
e^{i l \phi }
\left(
\begin{array}{c}
 \kappa K_l(\xi \rho ) \\
 0 \\
 0 \\
i e^{i \phi } \kappa \frac{\hbar c}{\mathcal{E}-U+m c^2} \{\frac{1}{2} \xi \left[-K_{l-1}(\xi \rho )-K_{l+1}(\xi \rho ) \right] -\frac{l}{\rho} K_{l}(\xi \rho ) \} \\
\end{array}
\right), & \rho >R, \\
\end{array}
\end{equation}
\end{widetext}
where $J_l$ and $K_l$ are the Bessel function and modified Bessel functions of order $l$, respectively.  The constant $\kappa$ measures the relative magnitude between the wavefunctions inside and outside the quantum well. 

We now apply the boundary condition to ensure all components of the spinor wavefunction is continuous at $\rho=R$:
\begin{eqnarray}\label{boundary}
&&\kappa  K_l(\xi  R)=J_l(\zeta  R)\nonumber \\
&&\kappa \{\frac{1}{2} \xi \left[-K_{l-1}(\xi R )-K_{l+1}(\xi R ) \right] -\frac{l}{R} K_{l}(\xi R ) \}= \nonumber \\
&&\frac{\mathcal{E}-U+mc^2}{\mathcal{E}+mc^2}\{\frac{1}{2} \zeta \left[J_{l-1}(\zeta R )-J_{l+1}(\zeta R ) \right] -\frac{l}{R} J_{l}(\zeta R)\}, \nonumber \\
\end{eqnarray}
from which the eigenenergies $\mathcal{E}_{ln}$ and constant $\kappa$ can be solved numerically. Here, $l$ and $n$ denote the azimuthal and radial quantum numbers, respectively. Numerical calculations can then be carried out to study the wave properties in all regions. 

The boundary conditions described in Eqs.~\ref{boundary} ensure the continuity of both the charge and current densities of the electron wave, while in the Schrödinger framework, only charge continuity is typically considered. The continuity of the spinor wavefunction thus preserves the Lorentz invariance of the electron throughout, except when interacting with a Lorentz-violating field, as previously studied in~\cite{PhysRevD.94.115020}. In that study, the eigenenergy levels—and consequently the spectra—of a Dirac electron in a cylindrical potential well were analyzed to investigate the effects of classical gravity within the quantum mechanical regime. However, the wave effects, including wave spin, were not explored, as it was stated on the third page of that paper that "we are not concerned with any global properties of the wave function in this work."

\section{\label{sec:Evanes}Evanescent electron wave}
To simplify the discussion on the evanescent electron wave, we choose the lowest azimuthal quantum number $l=0$ and the wavefunction becomes
\begin{equation}\label{tildepsil0}
\tilde{\psi }(\phi,\rho)=\Bigggg \{
\begin{array}{ll}
\left(
\begin{array}{c}
 J_0(\zeta \rho ) \\
 0 \\
 0 \\
-i e^{i \phi } \frac{\hbar c}{\mathcal{E}+m c^2}\frac{1}{2} \zeta J_1(\zeta \rho )\\
\end{array}
\right), & \rho \leq R; \\\\
\left(
\begin{array}{c}
 \kappa K_0(\xi \rho ) \\
 0 \\
 0 \\
-i \kappa e^{i \phi }\frac{\hbar c}{\mathcal{E}-U+m c^2} \frac{1}{2} \xi K_1(\xi \rho )\\
\end{array}
\right), & \rho >R. \\
\end{array}
\end{equation}

The boundary conditions become
\begin{eqnarray}\label{boundaryl=0}
\kappa  K_0(\xi  R)&=&J_0(\zeta  R)\nonumber \\
\kappa \xi K_1(\xi R )&=&\frac{\mathcal{E}-U+m c^2}{\mathcal{E}+m c^2} \zeta J_1(\zeta R ), \nonumber \\
\end{eqnarray}
which is combined to give the eigenvalue equation
\begin{equation}\label{boundaryl=0comb}
\xi J_0(\zeta  R) K_1(\xi R )=\frac{\mathcal{E}-U+m c^2}{\mathcal{E}+m c^2} \zeta K_0(\xi  R)J_1(\zeta R ),
\end{equation}
from which the eigenenergies $\mathcal{E}$ and constant $\kappa$ can be solved numerically. 

\begin{figure}
\includegraphics[width=0.45\textwidth]{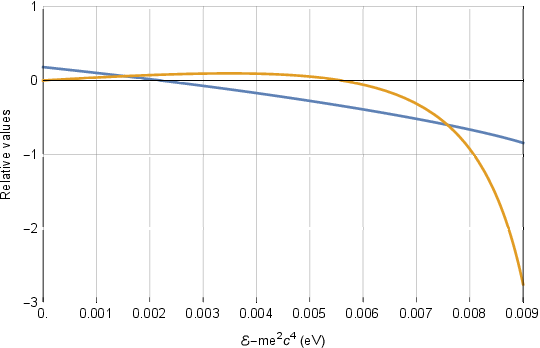}%
\caption{\label{fig:eigenU001}Numerical solution of the eigenenergies for the electron in a quantum well of $U=0.01~\text eV$ and $R=10~\text {nm}$. \\
The blue and orange curves representing $ \xi J_0(\zeta  R) K_1(\xi R )$ and $\frac{\mathcal{E}-U+m c^2}{\mathcal{E}+m c^2} \zeta K_0(\xi  R)J_1(\zeta R ) $ from Eq.~\ref{boundaryl=0comb} respectively intercept at eigenenergies $\mathcal{E}_{01}-m c^2=1.53 (meV)$ and $7.63 (meV)$ for the ground and excited state respectively.}
\end{figure}

We now conduct the numerical study by choosing a quantum well of radius $R=10~\text nm$ and a series of potentials $U=0.01, 0.1, 1, 10~\text eV$. For each potential, multiple eigenenergies can be found by solving the Eq.~\ref{boundaryl=0comb} numerically. As an example, for $U=0.01~\text eV$, only two eigenenergies are found within the quantum well, $\mathcal{E}_{01}-m c^2=1.53 (meV)$ and $\mathcal{E}_{02}-m c^2=7.63 (meV)$ that correspond to the ground state and first excited state, respectively. Fig.~\ref{fig:eigenU001} shows the two eigenenergy solutions as the inception points of functions $ \xi J_0(\zeta  R) K_1(\xi R )$ and $\frac{\mathcal{E}-U+m c^2}{\mathcal{E}+m c^2} \zeta K_0(\xi  R)J_1(\zeta R ) $ from Eq.~\ref{boundaryl=0comb}. 

We now calculate the ground state eigenenergies $\mathcal{E}_{01}-m c^2$ and $\kappa_{01}$ for all potentials, followed by the calculation of wave numbers $\zeta_{01}$ and $\xi_{01}$ with the help of Eqs.~\ref{zetaxi}. The results are listed in Table~\ref{tab:table1}. 
\begin{table}[b]
\caption{\label{tab:table1}%
Eigenenergy $\mathcal{E}_{01}$ and relative constant $\kappa_{01}$}
\begin{ruledtabular}
\begin{tabular}{ccccc}
\textrm{$U$ (eV)}&
\textrm{$\mathcal{E}_{01}-m c^2 (meV)$}&
\textrm{$\kappa_{01}$}&
\textrm{$\zeta_{01} \text (m^{-1})$}&
\textrm{$\xi_{01} \text (m^{-1})$}\\
\colrule
$0.01$ & $1.53$  & $44.1$ & $2.00\times 10^8$ & $4.71\times 10^8$\\
$0.10$ & $1.95$ & $2.23\times 10^6$ &$2.26\times 10^8$ & $1.60\times 10^9$ \\
$1.00$ & $2.12$ & $2.32\times 10^{21}$ &$2.36\times 10^8$ & $5.12\times 10^9$ \\
$10.0$ & $2.18$ & $1.75\times 10^{69}$ &$2.39\times 10^8$ & $1.62\times 10^{10}$\\
\end{tabular}
\end{ruledtabular}
\end{table}

\begin{figure}
\includegraphics[width=0.25\textwidth]{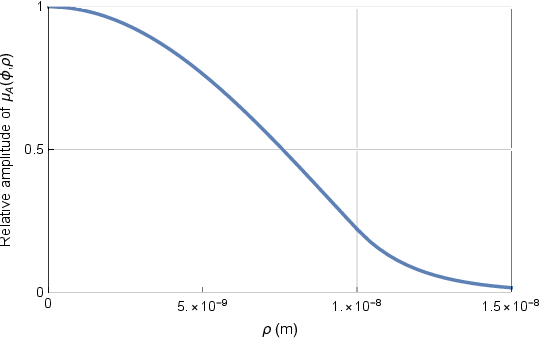}%
\includegraphics[width=0.25\textwidth]{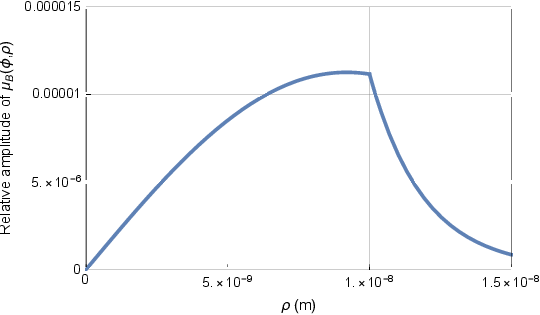}%
\caption{\label{fig:wavefunctionU001}Wavefunction plot for the Dirac electron in a quantum well of $U=0.01~\text{eV}$ and $R=10~\text {nm}$. \\
The large component $\mu_A(\phi ,\rho )$ (left) and small component $\mu_B(\phi ,\rho )$ (right) wavefunctions demonstrate that a substantial evanescent wave tunnels out of the quantum well, but satisfies wavefunction continuity at the boundary.}
\end{figure}

The numerical calculation shows that the electron wave tunnels out of the quantum well and behaves evanescent due to the characteristics of the modified Bessel functions.  Fig.~\ref{fig:wavefunctionU001} plots the wavefunctions inside and outside the well for the ground state of $U=0.01~\text eV$. It is shown that a substantial evanescent wave tunnels out of the quantum well, but satisfies wavefunction continuity at the boundary.

Table~\ref{tab:table1} shows that at higher potential, for example $U=10~\text eV$, the wave number $\zeta_{01}=2.39 \times 10^8$ already approaches the wave number for the infinite quantum well $\zeta_{01}^{\inf}=2.40 \times 10^8$, where zero wavefunction is assumed outside the well 
\begin{equation}
J_0(\zeta_{01}^{\inf}  R)=0.
\end{equation}

\begin{figure}
\includegraphics[width=0.25\textwidth]{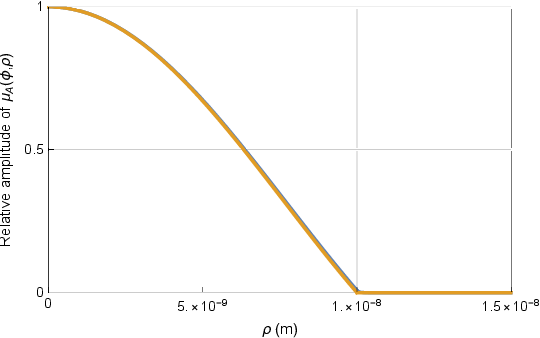}%
\includegraphics[width=0.25\textwidth]{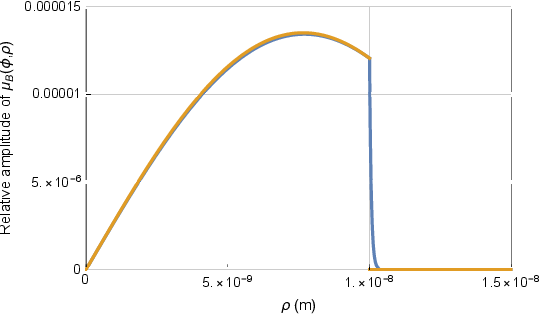}%
\caption{\label{fig:wavefunctionU10}Wavefunction plot for the Dirac electron in a quantum well of $U=10~\text{eV}$ and $R=10~\text {nm}$. \\
The large component $\mu_A(\phi ,\rho )$ (left) and small component $\mu_B(\phi ,\rho )$ (right) wavefunctions (blue)  inside the quantum well are nearly identical to the wavefunctions (orange) inside an infinite quantum well. The wavefunctions outside the quantum well are diminished but not exactly zero, especially at the boundary for $\mu_B(\phi ,\rho )$, so that the wavefunction continuity is always satisfied.}
\end{figure}

Therefore, the wavefunctions $\mu_A(\phi ,\rho )$ and $\mu_B(\phi ,\rho )$ inside the quantum well of relatively high potential $U=10~\text eV$ are nearly the same as the wavefunctions in the infinite quantum well, as shown in Fig.~\ref{fig:wavefunctionU10}. However, contrary to conventional assumptions, the wavefunctions outside the quantum well are not exactly zero; they are reduced but remain nonzero, particularly near the boundary for $\mu_B(\phi ,\rho )$. This nonzero presence near the boundary significantly contributes to the current density, as will be discussed in the next section. The wavefunctions remain substantial within a narrow region near the boundary, known as the skin depth, but they decay rapidly beyond this region. The skin depth becomes narrower as the potential increases, similar to the behavior of an optical field confined within a waveguide \cite{RefaieSR2023}.

\section{\label{sec:EvaneSpin}Evanescent electron wave spin}
We now investigate the spin nature of the evanescent electron wave that resides within the skin depth region outside the quantum well. 
Analytical expression of the current density is obtained by using the wavefunctions in Eq.~\ref{tildepsil0},
\begin{eqnarray}\label{currentl0}
j_{\rho }&=& -ec\psi ^{\dagger }\alpha _{\rho }\psi=0, \textrm everywhere; \nonumber \\
j_{\phi }&=& -ec\psi ^{\dagger }\alpha _{\phi }\psi=\biggl\{
\begin{array}{ll}
-\frac{e\hbar c^2}{\mathcal{E}+\text m c^2}\zeta
J_0(\zeta \rho )J_1(\zeta \rho ), & \rho \leq R \\\\
-\kappa ^2 \frac{e\hbar c^2}{\mathcal{E}-U+\text m c^2}\xi
K_0(\xi \rho )K_1(\xi \rho), & \rho >R, \\
\end{array}\nonumber \\
\end{eqnarray}
where $-e=-1.602 \times 10^{-19} \text C$ to represent the electron charge. 

\begin{figure}
\includegraphics[width=0.48\textwidth]{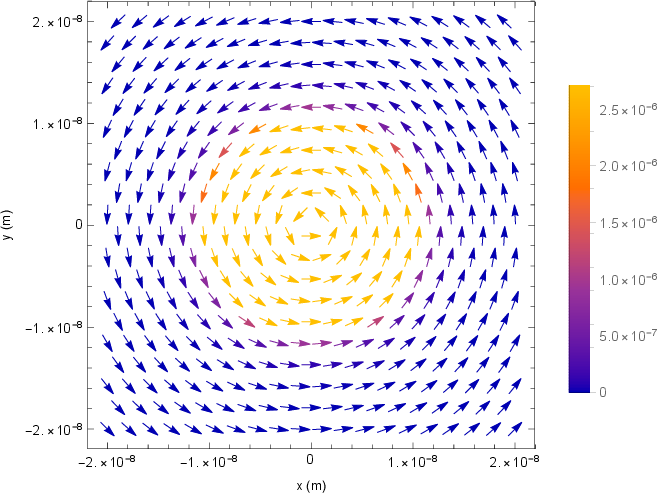}%
\caption{\label{fig:jvector}Vector plot of the current density for a spin-up ground state electron in a quantum well of $U=0.01~\text eV$ and $R=10~\text {nm}$. The evanescent wave spins concurrently with the wave inside the quantum well.}
\end{figure}

Eq.~\ref{currentl0} demonstrates that stable circulating current density exists both inside and outside the quantum well, as evidenced by the non-zero component $j_{\phi }$ in all regions and zero component $j_{\rho }$ everywhere. The evanescent electron wave is shown to spin concurrently with the electron wave inside the quantum well, as illustrated by the vector plot of Fig.~\ref{fig:jvector}. The current density continuity is observed at the boundary as a result of the wavefunction continuity at the boundary in Eqs.~\ref{tildepsi2} and \ref{tildepsil0}, which also ensures the charge density continuity. 

The above analysis underscores the wave spin picture, in which the spin is a wave property encoded in the entire Dirac field. In other words, the evanescent wave is an integral part of the entire electron wave that possesses the wave spin property. 

\section{\label{sec:probe}Evanescent electron wave spin sensing}
The discussion above raises questions about the security of spin quantum information and the possibility of novel evanescent electron wave spin sensing, which is analogous to evanescent optical wave sensing~\citep{Evanescent2019},~\cite{RefaieSR2023},~\cite{RefaieEPJST2023} that has already been employed in real applications. The evanescent optical wave is part of an eigen electromagnetic wave outside an optic waveguide or fibre that can interact with matters around the waveguide. The main optic wave confined inside the waveguide is perturbed but largely maintained. Similarly, the evanescent electron wave is a part of an eigen electron wave outside a quantum well that can interact with an electromagnetic wave around the quantum well. The optic and electronic evanescent wave detections share the same fundamental matter-field interaction: $\pmb{j} (\pmb r) \cdot \pmb {A}(\pmb r)$. The interaction is usually small due to the nature of evanescent waves; therefore, the main wave inside the confinement is only perturbed but not destroyed. Thus, evanescent wave sensing offers a unique detection scheme by enabling partial wave interactions without destroying the main wave inside the confinement, providing a different perspective on the quantum measurement problem~\cite{Leggett2005}.

However, the evanescent electron wave contains the spin property as described by the current density $\pmb{j} (r)$ in Eq.~\ref{currentl0}. When it interacts with the electromagnetic field $\pmb {A}(r)$, only partial spin participates in the process that could result in the fractional spin effects as discussed in the previous study~\citep{GaoQeios2023}. Such a partial spin concept conflicts with the particle spin picture, where a single particle electron possesses a unit spin and can only manifest the full spin effect during interaction. In the particle electron spin picture, the electron carrying the full spin tunnels out of the quantum well with a probability density given by the square of the wavefunction $\psi ^{\dagger }\psi$. If the spin is detected outside the quantum well, the spin information inside is destroyed since the electron that carries the full spin no longer exists within the quantum well.

\begin{figure}
\includegraphics[width=0.48\textwidth]{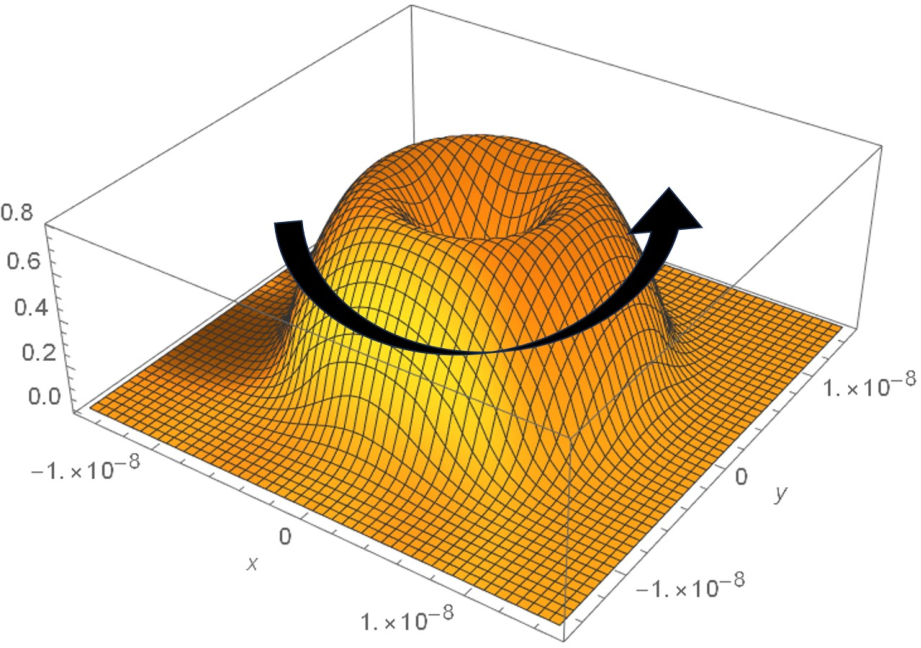} \\
\includegraphics[width=0.48\textwidth]{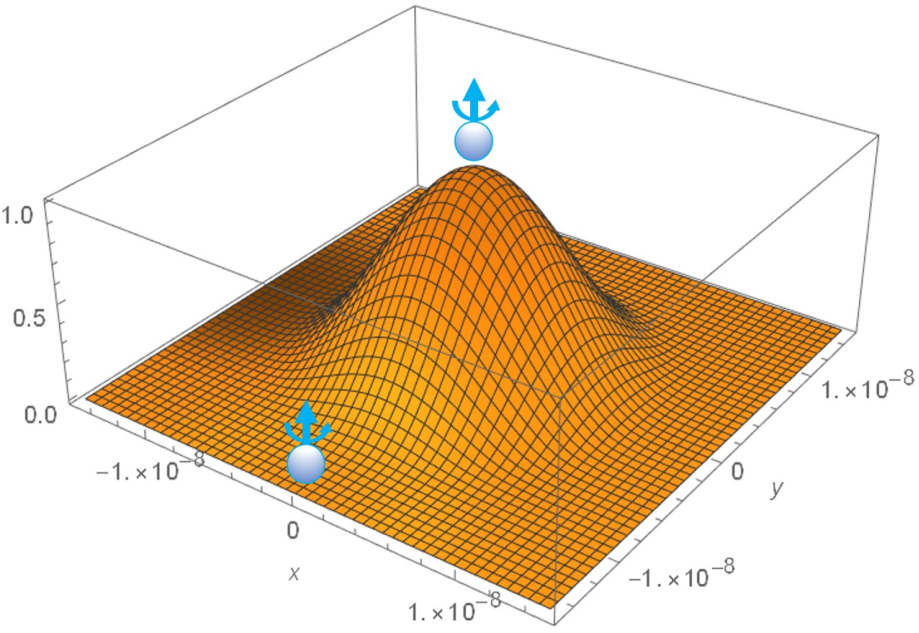}%
\caption{\label{fig:j03Dj3D} Conflicting views on the electron spin and tunnelling. \\
The upper figure illustrates the wave spin view that partial wave spin tunnels out of the quantum well with full certainty. The figure shows the three-dimensional distribution of the current density that spins as a whole in all regions for the spin-up electron in a finite quantum well of $R=10~\text {nm}$ and $U=0.01~\text eV$. \\
The lower figure illustrates the particle spin view that full particle spin tunnels out of the quantum well with partial certainty. The figure shows the probability density $\psi ^{\dagger }\psi$ of the particle electron of spin-up (represented by the ball and arrow) in the same quantum well. }
\end{figure}

The conflicting pictures are illustrated in Fig.~\ref{fig:j03Dj3D}, which comprises two figures for the electron in the same quantum well of $R=10~\text {nm}$ and $U=0.01~\text eV$. 

The lower figure illustrates the particle spin interpretation by displaying the probability density and particle electron spin. In this framework, based on wave-particle duality and the probabilistic interpretation of quantum mechanics, the entire electron spin can tunnel out of the quantum well, but only with a certain probability. If a measurement detects the spin outside the quantum well, it implies that no electron with spin remains inside the well.

In contrast, the upper figure depicts the wave spin interpretation, showing the spinning current density across all regions. The current density is similar to that found inside an infinite square well, as discussed previously~\citep{GaoJOPCO22}. In this view, a portion of the wave spin is present outside the quantum well with complete certainty. Here, it is possible to measure partial spin without disturbing the wave inside the well.

These conflicting views arise from different interpretations of the quantum mechanical wave, and these differences can be tested experimentally. In the conventional quantum mechanics framework, the microscopic world is fundamentally understood through the probabilistic wave-particle duality, where the wave is a statistical abstraction representing the probability distribution of the particle. In contrast, our interpretation posits that the electron is fully described by its wave properties, with the wave itself being the true and sole entity of the electron.  Since the wavefunction is a vector in the Hilbert space, the wave properties such as charge density $e\psi ^{\dagger }\psi$ and current density $ec\psi ^{\dagger }\alpha \psi$ are deterministic observables. Thus, the electron, in our view, is a concrete, deterministic entity fully characterized by Lorentz-invariant charge and current densities.

\section{\label{sec:Conclusion}Discussions}
In this work, we have demonstrated the existence of an evanescent electron wave outside both finite and infinite quantum wells. This result was obtained by solving the exact solutions of the Dirac equation for a cylindrical quantum well and applying the continuity conditions of the spinor wavefunction at the boundary. Additionally, we derived analytical expressions for the current density, which show that the evanescent wave exhibits a concurrent spin with the electron wave inside the quantum well.

The existence of a nonzero wavefunction outside an infinite quantum well implies that the electron wave cannot be strictly confined to a mathematical point. Additionally, the vortex topology demonstrated by the electron wave spin must also be preserved. These observations indicate that the singularities in mass density, and the consequent disruption of geodesics encountered in gravitational and cosmological contexts, may not accurately represent physical reality when the continuity of the Dirac wavefunction and the topological properties of wave spin are taken into account.

Moreover, the spin of the evanescent wave indicates that quantum information, or quantum entropy, is not entirely confined within the quantum well but can, in fact, extend beyond it. This insight opens up the possibility of accessing or detecting quantum spin information through the evanescent wave without causing a collapse of the entire spin state.

Our results underscore that the electron wave is a real, tangible entity with deterministic properties. This suggests that quantum processes or devices based on the manipulation and probing of electron waves are fundamentally deterministic, rather than probabilistic.

\section{\label{sec:Acknowledgement}Acknowledgement}
The authors wish to acknowledge the helpful discussions with the reviewers of this journal, as well as all the reviewers on the Qeios platform. Their insightful and constructive feedback significantly contributed to the improvement of this work.

\bibliography{Spin}

\end{document}